# Is a Single Photon Always Circularly Polarized? A Proposed Experiment Using a Superconducting Microcalorimeter Photon Detector

Alan M. Kadin, *Senior Member, IEEE*, and Steven B. Kaplan, *Senior Member, IEEE*

*Abstract*— A single photon is well known to have spin $S = \hbar$, which would correspond to circular polarization, and all quantum transitions with photon absorption or emission correspond to $\Delta S = \pm\hbar$. However, it is also widely believed that a single photon may be linearly polarized, which would correspond to a state with $S = 0$. Indeed, linearly polarized single photons are central to most quantum entanglement experiments. On the contrary, it has recently been suggested (based on a realistic spin-quantized wave picture of quantum states) that a linearly polarized photon state must be a superposition of a *pair* of circularly polarized photons, each with $S = \pm\hbar$. This question cannot be resolved using a conventional photon detector, which generally cannot distinguish one photon from two simultaneous photons. However, it can be addressed using a superconducting microcalorimeter detector with sub-eV energy resolution and high quantum efficiency (QE). A careful experiment demonstrating this photon pairing could place in question some of the paradoxical central foundations of modern quantum theory, including quantum entanglement and nonlocality.

## I. Photons and Polarization

A photon is a quantum of the electromagnetic field, with energy $E = \hbar\omega$, although there are still questions as to its proper physical representation [1,2]. According to the orthodox Copenhagen interpretation of quantum mechanics, a photon is either a point particle or a distributed wave, depending on the type of measurement. However, we would like to point out some implications of an unorthodox locally realistic wave picture of a photon [3], including a surprising prediction that can be tested experimentally using a superconducting photon detector. The focus here is on the polarization of a single photon. In particular, we will argue that a single photon must be a circularly polarized (CP) wave packet with distributed angular momentum totaling $S = \pm\hbar$ (see Fig. 1). Its spin is definite even if it has not been measured. In contrast, while a linearly polarized (LP) EM wave may be constructed as a vector superposition of two or more CP photons of opposite helicities, one cannot have a single LP photon. This is in contrast to the orthodox theory, in which a single photon may be prepared to have any polarization, including LP.

Experiments using LP single photons are ubiquitous in fundamental quantum optics [4], and it is universally believed that LP single photons have been routinely observed. However, we point out that most conventional photon detectors for visible light are event detectors that do not measure the absorbed energy [5,6,7]. Therefore, they cannot distinguish the absorption of a single photon from two simultaneous photons. In contrast, certain modern superconducting photon detectors are essentially microcalorimeters that measure the energy associated with a given absorption event [8,9,10,11,12,13,14,15]. To our knowledge, a careful energy-resolving experiment on purported LP single photons has not been reported. We suggest that such an experiment is necessary to confirm the existence of LP single photons. Furthermore, if it could be shown that all LP "single photons" are really photon *pairs*, then this would place into question the interpretation of an entire body of modern quantum experiments involving entanglement, nonlocality, and Bell's inequalities.

A classical TEM wavepacket is well known to carry energy and momentum distributed through its volume. One can define an energy density $\mathcal{E}$ and momentum density $\mathcal{P}$ from the

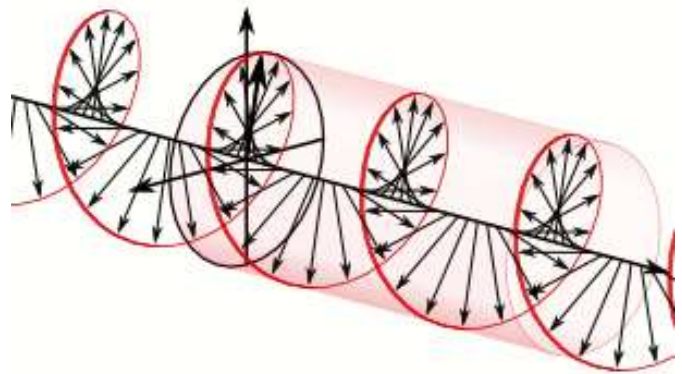

Fig. 1. Representation of a circularly polarized (CP) electromagnetic wave, with a rotating field-vector of fixed length (from [16]). A distributed wavepacket carries not only energy and momentum, but also angular momentum. If the total spin is $S = \hbar$, then $E = \hbar\omega$ follows from Maxwell's equations, with no other assumptions.

Submitted May 30, 2014.
A.M. Kadin is at 4 Mistflower Lane, Princeton Junction, NJ 08550 USA (e-mail: amkadin@alumni.princeton.edu).
S.B. Kaplan is at 1673 Summit Street, Yorktown Heights, NY 10598 USA (e-mail: stevebkaplan@optonline.net).

Poynting vector. It is perhaps less well known that the wavepacket can also carry distributed angular momentum density $\mathcal{L}$ associated with the rotation of the vector fields. While an LP wave has $\mathcal{L} = 0$, a CP wave (with a fixed-length field vector rotating with angular frequency ω) has $\mathcal{L} = \pm\mathcal{E}/\omega$ [3,17]. This is a standard problem in classical electromagnetics, for example in the classic text by Jackson [18].

The significance of this for quantum mechanics is that if one has a CP wavepacket with total energy $E = \hbar\omega$, then it also has a total angular momentum $L = \pm\hbar$. This, of course is the spin of the photon. This suggests that spin is not a mysterious intrinsic property of a point particle, but rather a globally conserved quantity of a real distributed wavepacket. This is a semiclassical picture of real waves in real space, with only the quantization of spin to turn this into a picture of discrete "particles" [3,19,20,21]. It is worth noting in this regard that angular momentum is one of the few physical quantities that is Lorentz invariant, so that spin $\hbar$ is the same for all inertial reference frames, even while the photon is red-shifted or blue-shifted.

Within this spin-quantized wave picture, multi-photon fields are wavepackets that are simply the vector sum of $n_R$ right CP photons and $n_L$ left CP photons, for a total angular momentum $L = (n_R - n_L)\hbar$, where $n_R$ and $n_L$ are non-negative integers. So the permitted states in a multi-photon field are represented by the lattice points in Fig. 2. This represents a two-dimensional Hilbert space, where the single CP photons represent the natural basis, and only integer linear combinations are permitted. This compares to a classical Hilbert space, where any linear combination would be permitted. Note that while LP composite states are allowed (along the diagonal), these must be composed of matched *pairs* of left CP and right CP photons. There are no LP single photons in this picture. Furthermore, there is no intrinsic quantum uncertainty; the spin of a given photon state is definite, even if it has not been measured.

This should also be contrasted with the orthodox quantum theory [22], in which a single photon can have any polarization, corresponding to a (non-integer) linear combination of CP photon states along the unit circle in Fig. 2. Furthermore in the orthodox theory, a multi-photon state is *NOT* a simple linear combination of single photon states (as in Fig. 2), but rather an entangled linear combination of *products* of single-photon states [23]. While it would be of great interest to provide an alternative explanation for experiments showing quantum entanglement, the present paper focuses on a simpler problem: can a single photon be measured to be linearly polarized?

How are single photons produced in experiments? There are essentially two approaches [6]. In one, a multi-photon beam (for example, from a laser) is highly attenuated until the detector observes a low rate of pulses. But some detectors may exhibit a background dark signal in the absence of an absorbed photon. Alternatively, one can use a source that generates simultaneous diverging photon pairs, and use the first photon of the pair as a coincidence trigger for the second photon. This is called a source of "heralded" single photons [6]. In either case, it may not be completely certain that a detected "single photon" is really just one. This is because photons in a source are always subject to stimulated emission, and can be highly correlated both in space and time.

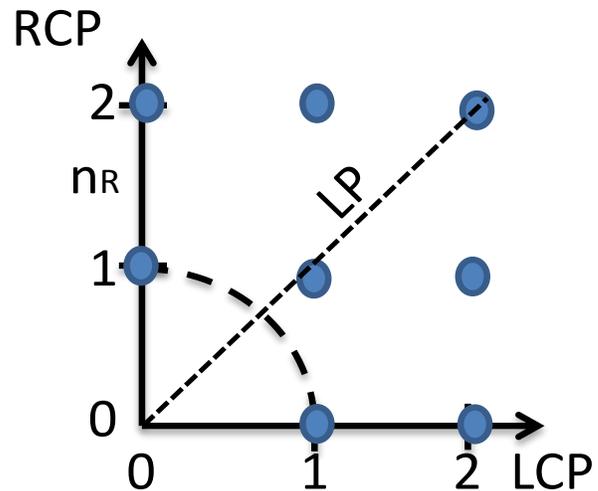

Fig. 2. Allowed states for photon field, as integer linear combinations of right circularly polarized (RCP) and left circularly polarized (LCP) single photons, according to the spin-quantized wave picture [3]. Linearly polarized states are along the diagonal, corresponding to even numbers of photons with total spin $S = 0$. In contrast, in the orthodox theory, a single photon can be any state along the unit circle, whereas multiphoton states are based on product states of single photons, not simple sums.

How is a linearly polarized single photon produced in experiments? Classically, a linear polarizer is a device that passes the component of electromagnetic radiation with electric field in, say, the x-direction, and absorbs the component vibrating in the y-direction. If one takes a beam of unpolarized single photons, and directs it to a linear polarizer, according to the orthodox theory, half of the photons will pass through as linearly polarized single photons polarized in the x-direction, and the other half of the photons will be absorbed. This is a quantum measurement and a random statistical process [22]. On the contrary, in the realistic wave picture presented here, a single photon is always CP, so that it will always get absorbed in a linear polarizer. On the other hand, a field of two photons that is linearly polarized in the x-direction will pass through the polarizer unchanged.

## II. SUPERCONDUCTING PHOTON DETECTORS

A photon detector is a device that absorbs a photon and provides an electrical pulse. In principle, one could directly detect the photoelectron emission from a single photon absorption event, but this is generally too weak for visible or infrared photons. Far more common is an avalanche detector, where a single photoelectron generates a cascade of amplification, leading to a much larger current pulse $Ne$, where $N$ may be thousands or more [7]. This is the case, for example, with a photomultiplier tube or a semiconductor avalanche photodiode. This is effectively similar to a superconducting nanowire single photon detector (SNSPD, [24]), where absorption of a single photon leads to an instability in a current-carrying superconducting nanowire, creating a normal hotspot and giving a much larger signal than that due to the single initial photoelectron. The problem with an avalanche photon detector (of any technology) is that it cannot accurately distinguish one absorbed photon from two such photons absorbed at (nearly) the same time. The avalanche process is highly nonlinear, and generally insensitive to the initial deposited energy.

What is needed to properly distinguish one from two photons is a detector that accurately measures the absorbed energy. This can be achieved for x-ray photons using semiconductor detectors [25], where a single x-ray photon creates a high-energy (~ keV) photoelectron that rapidly distributes its energy (without amplification) among ~ 1000 low-energy secondary electrons ~ eV, and the current pulse from the secondary electrons is collected before these relax back to their ground state. The number of such secondary electrons is proportional to the energy of the initial x-ray photon. For visible photons ~ 1 ev, however, the excitation energy of secondary electrons in semiconductors is too large to provide sub-eV energy resolution. In contrast, superconductors offer enhanced energy resolution – the energy of excited quasiparticles is ~1 meV for low-temperature superconductors, and the thermal smearing is even smaller. So a ~1eV photon can rapidly lead to ~ 1000 1-meV excitations, providing sufficient energy resolution. A device with energy sensitivity on such small scales is known generically as a microcalorimeter. If this is combined with an absorber with high QE, one has an energy-sensitive single-photon detector.

Several types of superconducting devices can operate as microcalorimeter photon detectors [26], including Transition Edge Sensors (TES), Superconducting Tunnel Junction Detectors (STJ), and Kinetic Inductance Detectors (KID). Similar structures can alternatively function as microbolometers for lower-energy photons (farther into the infrared and microwave range), where the individual photon energies are smaller than the energy resolution [27]. For single-photon detectors with fine energy resolution, much of the recent attention has focused on TES devices that consist of a small, thin superconducting film biased at the resistive transition, typically well below 1 K. An absorbed photon increases the effective temperature, thus increasing the film resistance. Some TES devices have been demonstrated with energy sensitivity down to 0.1 eV and QE better than 90%, with very low dark signals [9,5,28,29].

Several experiments have demonstrated the ability of these TES detectors to count the number of photons in an attenuated, unpolarized beam from a pulsed laser [8,9,10]. For example, the beam may be attenuated so that the average number of photons per pulse is about 1. But this is a statistical average, so the actual number of photons in a given pulse may be 0, 1, 2, or even 3 or 4 with reduced probability (with a Poisson distribution). This was carried out for an unpolarized beam. A histogram of the pulse counts from [9] is shown in Fig. 3, confirming this. Here the photon energy was 0.8 eV,

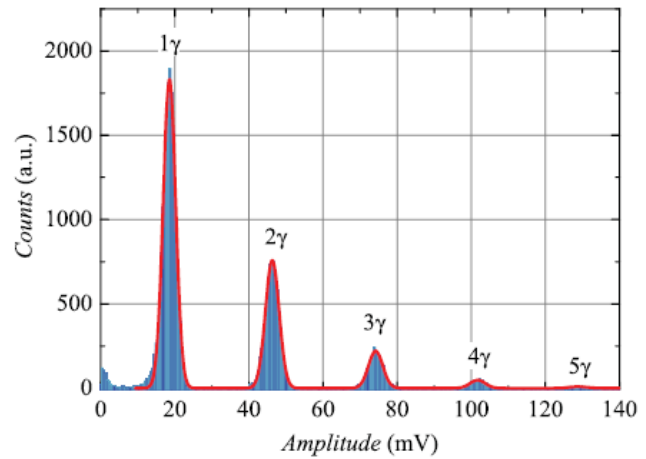

Fig. 3. Photon number histogram of beam of attenuated unpolarized pulses, each with an average of about 1 photon per pulse (from [9]).

and the energy resolution 0.1 eV.

## III. PROPOSED EXPERIMENT

How would the results of Fig. 3 be modified if the incident unpolarized beam were passed through a linear polarizer before reaching the detector? According to the orthodox theory, the result would be very similar: a Poisson distribution of single photons. On the contrary, according to the spin-quantized wave picture, if these laser pulses are linearly polarized coherent electromagnetic wavepackets, then they must consist of *pairs* of CP photons. In that case, a photon counting detector (with QE = 1) should see only even numbers of photons. The extra single photons will have been absorbed by the polarizer. This is the proposed experiment, as shown in the block diagram in Fig. 4. This shows essentially the same experiment repeated with both unpolarized and polarized laser pulses.

Note that this observation of the extinction of odd photon numbers requires a photon detector with a very high QE. A photon pair may appear to be a single photon if QE is low. A QE <1 will result in depressed (but nonzero) peaks for odd integers. Another variant of this proposed experiment would be to use a circularly polarized source rather than a linearly polarized source. This would be expected to yield both even and odd photon numbers, in contrast to linearly polarized pulses.

## IV. DISCUSSION AND CONCLUSIONS

This spin-quantized wave picture for photons is strikingly different from the orthodox picture for correlated states in classic problems such as the Einstein-Podolsky-Rosen (EPR) paradox [30], which in turn led to the recognition of quantum entanglement. Consider, for example, a spin-zero initial state that decays to two photons. This occurs, for example, in positron annihilation to two gamma rays, but a similar decay may occur in atomic systems as well. In the orthodox theory, the resulting photons are initially undefined in direction, momentum, and spin, but they are in a two-body correlated state (even if they are far apart), which by construction must be entangled. When at least one of these photons is measured, the two-body correlated state immediately decouples (decoherence, sometimes called the "collapse of the wave function"), leading to two separated photons with opposite parameters. For example, if linear polarization is measured, one photon will be vertically polarized while the other is horizontally polarized. Numerous experiments have confirmed this [4], and the analyses (in terms of the Bell inequalities [31]) seem to be incompatible with "local hidden variables", in which the two photons could have definite values prior to measurement. However, these analyses assume that the detected photon is a single photon, although as noted above, earlier generations of detectors could not generally distinguish a single photon from two simultaneous photons. We suggest that this may provide a "loophole" for locally realistic quantum states, and that the experiments should be redone using a calibrated superconducting microcalorimeter photodetector.

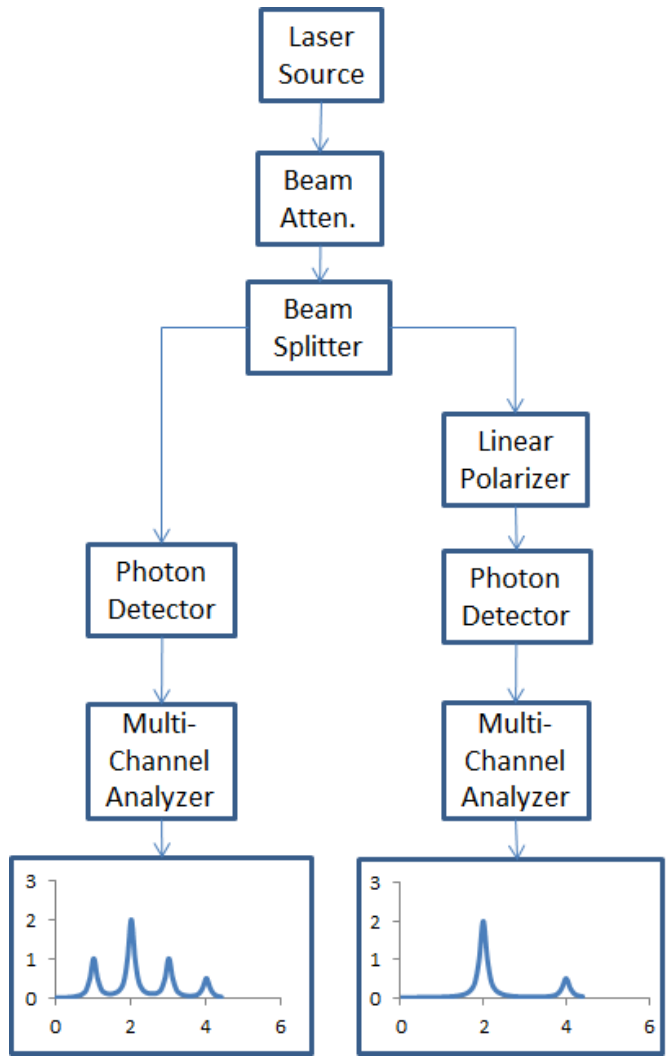

Fig. 4. Proposed experiment using superconducting photodetector with and without linear polarizer. With the polarizer in place, the peaks corresponding to odd photon numbers should disappear (according to the spin-quantized wave picture [3]), leaving only the even photon peaks corresponding to linearly polarized photon pairs.

In the spin-quantized wave picture, the two photons are always wave packets with definite spin moving in opposite directions (see Fig. 5). Once they separate, they may remain correlated, but they are not coupled, so that they are not entangled. Specifically, they may both be left CP, or both right CP, which corresponds to opposite spin. Single-photon states are never linearly polarized (which would not quantize spin), but one might have LP two-photon states. It is not yet clear that such photon pairs could reproduce the results of the entanglement experiments, but this should be analyzed in greater detail.

A further distinction between the orthodox theory and the present realistic picture is the mathematical formalism of multi-photon states. The orthodox theory regards a two-photon state as a symmetrized product of two single-photon states: $\Psi_{tot} = \Psi_{1A} \times \Psi_{2B} + \Psi_{1B} \times \Psi_{2A}$, where 1 and 2 represent the two photons, and A and B represent complementary properties of the photons (such as vertical and horizontal polarization). Such a construction is intrinsically entangled, and incompatible with local reality [32]. In contrast, in the spin-quantized wave picture, $\Psi_{tot} = \Psi_A + \Psi_B$, simply a real-space sum of two wave packets (as in Fig. 2), the same as with classical waves. Once these separate in space, they are no longer linked in any way, so no instantaneous collapse of the wave function is necessary.

Product states and entanglement are also central to quantum information theory [3232], which informs the developing applications of quantum communication and quantum computing. If those mathematical constructions are not present in real quantum states, that would place the entire foundation of these applications in jeopardy. Careful measurements of polarized photons should be carried out with this in mind.

In conclusion, we have suggested an experiment using superconducting energy-sensitive single-photon detectors to carefully observe the presence or absence of linearly polarized single photons. Such LP single photons have been claimed in many experiments, but not to our knowledge using detectors that could distinguish one from two simultaneous photons. If would-be LP "single photons" turn out to be photon *pairs* instead, that would indicate that the orthodox quantum theory has serious foundational problems, and that a locally realistic quantum picture, without indeterminacy or entanglement, should be seriously reconsidered.

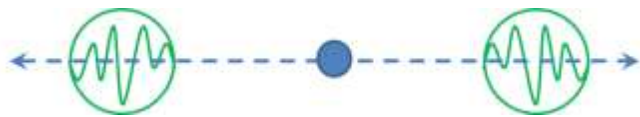

Fig. 5. Representation of correlated two-photon state following decay of a spin-zero precursor. In the orthodox quantum theory, these photons have undefined individual properties and are entangled until a measurement. In the spin-quantized wave picture, once separated in space, these are two real wave packets with definite properties, which are not coupled to one another.

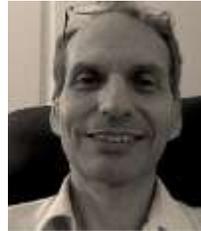
**Alan M. Kadin** (M'87–SM'14) received the B.A. in physics from Princeton University in 1974, and the M.A. and Ph.D. in physics from Harvard University in 1975 and 1979, with a thesis on superconducting devices.

He was a Postdoctoral Research Associate at Stony Brook University, New York from 1979-1981 and at University of Minnesota, Minneapolis from 1981-1983. He was then a Research Scientist at Energy Conversion Devices and its subsidiary Ovonic Synthetic Materials Company, in Troy, Michigan, from 1983-1987. He then served as an Associate Professor of Electrical and Computer Engineering at the University of Rochester (New York) from 1987-2000. From 2000-2005 he was a Senior Scientist at Hypres, Inc., Elmsford, NY. Since 2005 Dr. Kadin has been a Technical Consultant based in Princeton Junction, New Jersey. Concurrently in 2012 he was an Adjunct Professor of Electrical and Computer Engineering at the College of New Jersey, Ewing, NJ. He is the author of the textbook *Introduction to Superconducting Circuits* (Wiley, New York, 1999), and over 100 publications and 10 patents.

Dr. Kadin is a longtime specialist in superconducting circuits, devices, and materials, with a continuing interest in the foundations of physical theories. He is a member of the American Physical Society. He is currently participating in the IEEE Rebooting Computing Working Group.

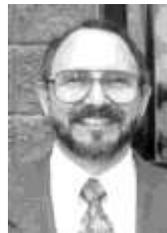
**Steven B. Kaplan** (M'96–SM'05) received the B.A. in physics from Princeton University in 1971, and the Ph.D. in physics from the University of Pennsylvania in 1977, with a thesis on quasiparticle lifetimes and nonequilibrium superconducting states. He was a National Research Council Postdoctoral Fellow at the National Bureau of Standards (now

NIST), Boulder CO from 1977-1979. He was then a Research Staff Member at the IBM Thomas J. Watson Research Center, Yorktown Heights NY from 1979-1990. He then taught Physics, Mathematics and Statistics at Dutchess Community College, Poughkeepsie NY from 1990-1991. Dr. Kaplan then spent 20 years at HYPRES, Inc. Elmsford NY, first as a Research Staff Member, then as Senior Staff Member and Senior Scientist from 1991 to 2011. Since 2011, Dr. Kaplan has been a Technical Consultant based in Yorktown Heights, New York. Most recently, he has worked on neural superconducting circuits. He has one patent, and is the author of more than 40 publications, one of which has accumulated 500 academic citations.

Dr. Kaplan is a longtime specialist in superconducting circuits, devices, and nonequilibrium effects in superconductors. He also worked in mesoscopic physics and 2-D effects in conductors. His most recent interests lie in the theory of superconductivity and quantum mechanics. He is a member of the American Physical Society.